# Hysteretic Electronic Phase Transitions in Correlated Charge-Density-Wave State of 1$T$-TaS$_2$

Yanyan Geng[1], Le Lei[1,2], Haoyu Dong[1], Jianfeng Guo[1], Shuo Mi[1], Yan Li[1], Li Huang[2], Fei Pang[1], Rui Xu[1], Weichang Zhou[3], Zheng Liu[4], Wei Ji[1,*] and Zhihai Cheng[1,*]

[1]*Beijing Key Laboratory of Optoelectronic Functional Materials & Micro-nano Devices, Department of Physics, Renmin University of China, Beijing 100872, People's Republic of China*

[2]*Beijing National Laboratory for Condensed Matter Physics, Institute of Physics, Chinese Academy of Sciences, Beijing 100190, China*

[3]*Key Laboratory of Low-dimensional Quantum Structures and Quantum Control of Ministry of Education, College of Physics and Information Science, Hunan Normal University, Changsha 410081, People's Republic of China*

[4]*Institue for Advanced Study, Tsinghua University, Beijing 100084, China*

**Abstract:** Recently, many exotic electronic states, such as quantum spin liquid (QSL) and superconductivity (SC), have been extensively discovered and introduced in layered transition metal dichalcogenides 1$T$-TaS$_2$ by controlling their complex correlated charge-density-wave (CDW) states. However, few studies have focused on its hysteretic electronic phase transitions based on the in-depth discussion of the delicate interplay among temperature-dependent electronic interactions. Here, we reported a sequence of spatial electronic phase transitions in the hysteresis temperature range of 1$T$-TaS$_2$ via variable-temperature scanning tunneling microscopy (VT-STM). The emergence, evolution, coexistence, and separation of diverse novel electronic states within the commensurate CDW/triclinic CDW (CCDW/TCDW) phase are investigated in detail through the warming/cooling process. These novel emergent electronic states can be attributed to the delicate temperature-dependent competition and/or cooperation of interlayer interactions, intralayer electron-electron correlation, and electron-phonon (*e-ph*) coupling of 1$T$-TaS$_2$. Our results not only provide a novel insight to understand the hysteretic electronic phase transitions of correlated CDW state, but also pave a way to realize more exotic quantum states by accurately and effectively controlling various interactions in correlated materials.

*Corresponding author. E-mail address: zhihaicheng@ruc.edu.cn, wji@ruc.edu.cn

**Introduction**

Low-dimensional materials can exhibit various intriguing electronic ordering states, including charge-density-wave (CDW), spin-density-wave (SDW), superconductivity (SC), electronic nematicity, and so on [1–7]. As an archetype charge-ordered state, the CDW state normally consists of a periodic modulation of the electronic charge density accompanied by a periodic distortion of the atomic lattice [8, 9]. The most commonly considered mechanisms for the formation of conventional CDW states are related to a Fermi surface nesting and electron-phonon (*e-ph*) coupling [10–12]. A series of complicated CDW states have triggered off tremendous research interest in two-dimensional transition metal dichalcogenide (TMD) materials, such as 2$H$-NbSe$_2$ (3 × 3 CDW) [13, 14], 2$H$-TaSe$_2$ (3 × 3 CDW) [13], 1$T$-TiSe$_2$ (2 × 2 CDW) [15], 1$T$-TaSe$_2$ ($\sqrt{13} \times \sqrt{13}$ CDW) [16, 17], and monolayer VSe$_2$ ($\sqrt{7} \times \sqrt{3}$ CDW) [13]. More exotic electronic states, including SC [18], Anderson insulator [19], and normal metal [19] can be further induced in these TMD materials by effectively controlling the CDW states through external pressure [20], element doping [21], interlayer intercalation [22], etc.

Bulk 1$T$-TaS$_2$ is a well-known CDW material with a sequence of phase transitions with decreasing temperatures: from a metallic incommensurate CDW (ICCDW) phase through a nearly commensurate CDW (NCCDW) phase to a correlated commensurate CDW (CCDW) phase [23–25]. The correlated CCDW phase has attracted extensive attention due to the complex electron correlations effect and possible existence of a quantum spin liquid (QSL) state with localized electron charge and spin in the triangular CCDW lattice. It was previously assumed that the ground CCDW state of 1$T$-TaS$_2$ is driven by the single-band Mott-Hubbard mechanism [28], whereas the stacking-dependent interlayer interaction has recently been emphasized as an indispensable role to understand the insulating mechanism of 1$T$-TaS$_2$. Recent STM and angle-resolved photoemission (ARPES) measurements gave some robust evidence of dimerization along the *c*-axis, and simplified the insulating state into a trivial band insulator [29–32].

At low temperatures, the precise site-specific scanning tunneling spectroscopy (STS) measurements [29, 32, 33] have been used to directly distinguish the Mott- or band-insulating regions on the surface of 1$T$-TaS$_2$. At high temperatures, the temperature-dependent ARPES measurements indicated a distinctive electronic phase transition from the band insulator with interlayer dimerization gradually to the Mott insulator without interlayer stacking order within the CCDW–TCDW (triclinic CDW) hysteresis temperature range [34]. Such electronic phase transition with the decoupling of the interlayer stacking order can not only be triggered by changes in temperature but also by means of strain/pressure, interlayer intercalation, and element substitution [20–22]. To deeply understand and effectively control this electronic phase transition of $1T$-TaS$_2$, it is necessary to study the competition of stacking-dependent interlayer interactions and intralayer interactions in detail. And for any tunable interactions of $1T$-TaS$_2$, the evolution of the electronic phase transition itself becomes of great importance. Making a deep investigation of hysteric CCDW–TCDW phase transition and observing the spatial variation of the correlated CDW states of $1T$-TaS$_2$ may contribute to a further in-depth understanding of the temperature-dependent various interactions in the correlated CDW states.

In this paper, the CCDW–TCDW hysteretic phase transition of $1T$-TaS$_2$ were investigated in real space by variable-temperature scanning tunneling microscopy (VT-STM). Several successive stages, including the emergence, evolution, coexistence, and separation of various exotic electronic states at multi-scale (from the atomic-scale to mesoscopic-scale) were observed during the warming/cooling CCDW–TCDW phase transitions. We further discussed the effect of the interplay between interlayer and intralayer interactions on the exotic electronic states within the CCDW/TCDW phase of $1T$-TaS$_2$. Our results indicated that the competition between interlayer interactions and intralayer electron-electron correlation is responsible for the novel emergent electronic states in the CCDW phase, and the cooperation of intralayer electron-electron correlation and *e-ph* coupling induces the emergence of the novel emergent electronic states in the TCDW phase.

## Results

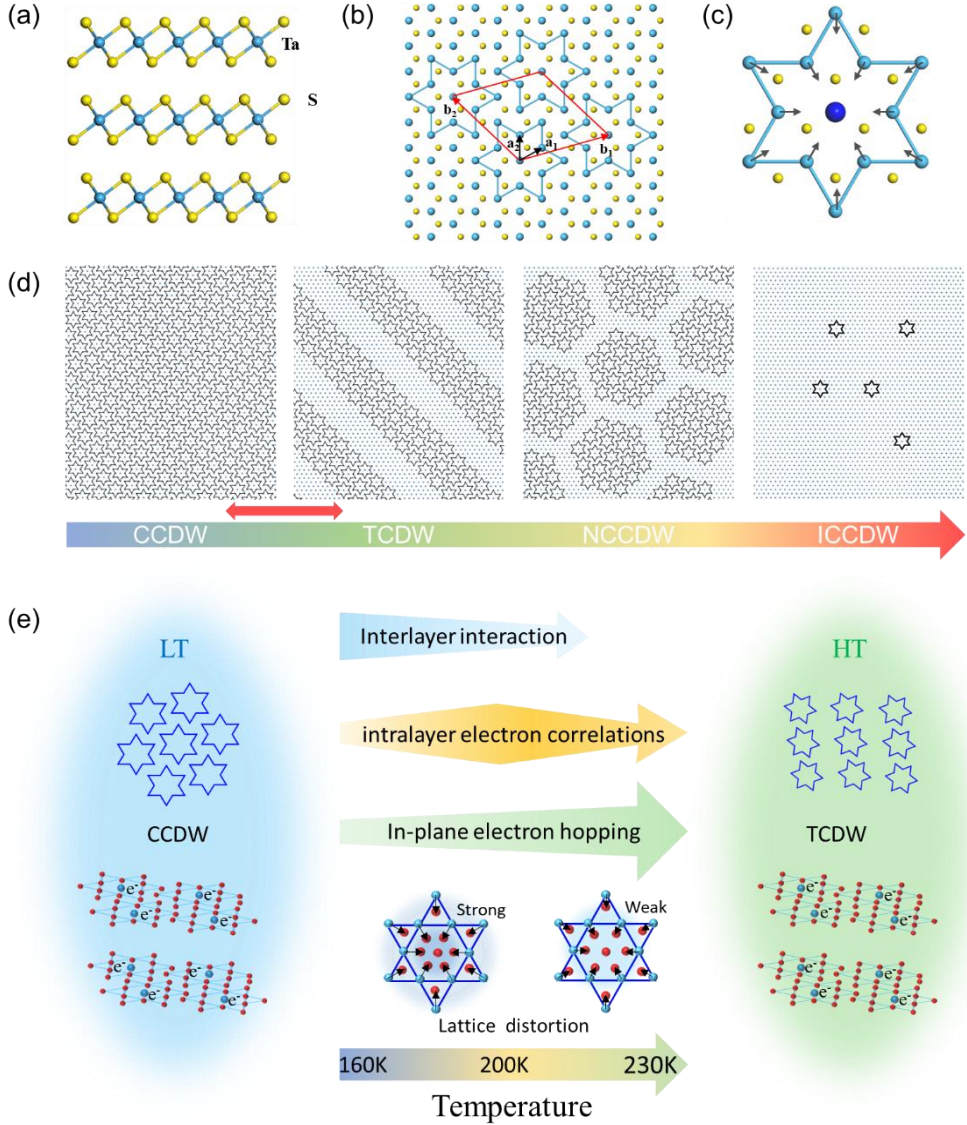

**FIG. 1. Atomic structure and correlated CDW states of 1*T*-TaS$_2$.** (a) Atomic crystal structure of the bulk 1*T*-TaS$_2$. (b) Schematic of the ground CCDW state of monolayer 1*T*-TaS$_2$ with a √13 × √13 superlattice. The ***b***$_1$ and ***b***$_2$ are the basis vectors of the CCDW superlattice, whereas ***a***$_1$ and ***a***$_2$ are the basis vectors of the underlying atomic lattice. (c) Schematic of the structure of the star-of-David (SD) unit in the CCDW state. The grey arrows indicate that the 12 surrounding Ta atoms shrink to the central Ta (blue) atom. (d) Schematics of the sequential CCDW, TCDW, NCCDW and ICCDW phases during the warming process. The SD clusters are marked with dark stars. (e) Schematic of the delicate interplay among the complex interlayer and intralayer interactions during the warming CCDW–TCDW phase transition of 1*T*-TaS$_2$.

As shown in Fig. 1(a), 1*T*-TaS$_2$ exhibits a CdI$_2$-type layered crystal structure with Ta atoms octahedrally coordinated by S atoms. At low temperatures, strong *e-ph* coupling induced periodic lattice distortion makes a perfect √13 × √13 superlattice in the ground

CCDW state [11, 44], in which every 13 Ta atoms are distorted to form the star-of-David (SD) clusters, as shown in Figs. 1(b) and 1(c). Each SD cluster contains 13 Ta 5$d$ electrons, the 12 electrons of the outer Ta atoms pair and form 6 occupied insulating bands and leave one unpaired electron of the central Ta atom in a half-filled metallic band [38, 39]. The strong intralayer electron-electron correlation introduces a Mott gap of about 0.2–0.4 eV at low temperatures, forming a correlated Mott insulator without considering the interlayer interaction [29, 33]. Recently, the important role of the interlayer stacking order has been discovered, and the ground CCDW state of 1$T$-TaS$_2$ can be described as a trivial band insulator/conventional Mott insulator on the termination of $T_A/T_c$ stacking at low temperatures [29, 32]. Intriguingly, these stacking-sensitive interlayer interactions not only play a dominant role in the ground CCDW state, but also have a significant effect on the electronic phase transitions in the correlated CDW state of 1$T$-TaS$_2$, particularly the CCDW–TCDW phase transition.

As shown in Fig. 1(d), bulk 1$T$-TaS$_2$ undergoes a series of CDW transitions with prominent hysteretic behavior at ~200 K. During the cooling process, 1$T$-TaS$_2$ directly transforms from the NCCDW phase into the CCDW phase at ~160 K with persistent three-fold rotation symmetry. However, during the warming process, 1$T$-TaS$_2$ undergoes a relatively complex CDW transition from the three-fold symmetric CCDW phase to the stripe-like nematic TCDW phase at ~220 K. Subsequently, the 1$T$-TaS$_2$ enters an NCCDW phase at ~280 K with hexagonal CCDW domains in discrete distribution. These localized CCDW domains shrink upon warming and finally disappear at ~350 K, when the system enters a metallic ICCDW phase. In the above phase transitions, although most previous studies have been extensively focused on the CCDW–NCCDW phase transition [40–42], we mainly focus on the spatial CCDW–TCDW phase transition owing to the more intense competition of various interactions of 1$T$-TaS$_2$.

The schematic in Fig. 1(e) illustrates the delicate interplay among the interlayer and intralayer interactions during the CCDW–TCDW phase transition of 1$T$-TaS$_2$. The low-temperature CCDW state is primarily governed by strong interlayer interactions with the stable stacking order rather than intralayer electron-electron correlation [29–31, 47]. With an increase in temperature, the interlayer interactions gradually decrease from ~160 K with the expansion of *c*-axis [33] and the stacking order is rather random with many unstable stacking interfaces [30]. Meanwhile, the intralayer electron-electron correlation gradually has a distinctive effect on the correlated CDW state. At temperatures above ~200 K, the interlayer stacking order completely vanishes and the intralayer electron-electron correlation dominant [34]. In addition, the warming process is also accompanied by the weakening of *e-ph* coupling and relative enhancement of in-plane electron hopping [22, 44]. At temperatures above ~230 K, the delicate interplay between the *e-ph* coupling and electron-electron correlation induces the emergence of the TCDW phase. This temperature-dependent competition between the interlayer and intralayer interactions can induce diverse spatially distributed exotic electronic states in the CCDW–TCDW phase transition range of 1*T*-TaS$_2$.

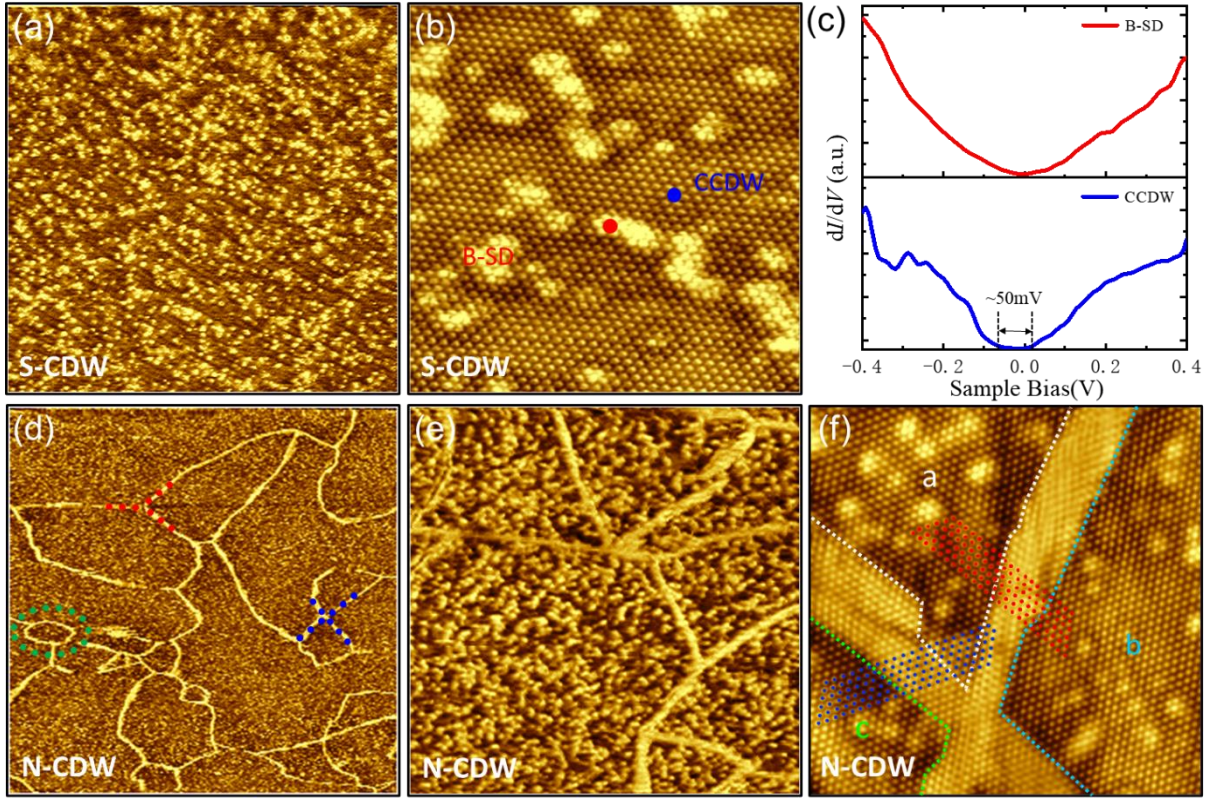

**FIG. 2. Emergent spotty-CDW (S-CDW) and network-CDW (N-CDW) electronic states within the CCDW phase during the warming process.** (a) Large-scale STM image of the CCDW phase with the emergent distributed bright clusters, which is called the S-CDW state. (b) High-resolution STM image of the S-CDW state. (c) Site-specific STM d$I$/d$V$ spectroscopy performed at the bright clusters and CCDW positions marked in (b). The small gap ~50 mV in the CCDW region is clearly identified, while the bright clusters exhibit a V-shaped feature with no gap. (d) Mesoscopic-scale STM image of the CCDW phase with the emergent distributed bright clusters and bright stripe network, which is called the N-CDW state. The network consists of bright stripes connected via the "Y-shape" (red), "X-shape" (blue), and "loop-shape" (green). (e) Large-scale STM image of the N-CDW state. (f) High-resolution STM image of the N-CDW state. The different S-CDW domains (marked by a, b, and c) are separated by the bright stripe network. Imaging parameters: 0.25 × 0.25 um$^2$ (a); 50 × 50 nm$^2$ (b); 1.5 × 1.5 um$^2$ (d); 0.15 × 0.15 um$^2$ (e); 75 × 75 nm$^2$ (f). $V_b$ = 200 mV, $I_t$ = 100 pA (a–f).

At temperatures above ~160 K, randomly distributed bright clusters first emerge within the CCDW phase, as shown in Figs. 2(a) and 2(b), which is referred to as the spotty-CDW (S-CDW) state at here. These bright clusters normally consist of approximately 5-8 bright SD (B-SD) and account for ~20% of the entire CCDW phase. The apparent heights of the B-SDs are ~ 50 pm higher than that of the remaining pristine SDs of the CCDW state [Fig. S1]. The local electronic states of the bright clusters (red dot) and pristine CCDW regions (blue dot) are further determined through the site-specific STS spectroscopy measurements, as shown in Fig. 2(c). For the pristine CCDW regions, the small gap of ~50 mV can be clearly identified even at the temperature of ~160 K, whereas the bright clusters exhibit a V-shaped feature with no gap. It is clear that the small gap of the pristine CCDW regions may originates from the operative intralayer electron-electron correlation with diminishing interlayer interactions. These emerging metallic bright clusters can be considered as the release regions for strong intralayer electron repulsion where the electrons are delocalized.

Figures 2(d), 2(e), and 2(f) show a further emergent bright stripe network within the S-CDW state at temperatures above ~180 K. The mesoscopic bright stripe network consists of bright clusters and the long bright stripes connected via the "Y-shape," "X-shape," and "loop-shape," as shown in Fig. 2(d), which is called the network-CDW (N-CDW) state. In the N-CDW state, S-CDW domains with various sizes and shapes are separated by the bright stripe network, as shown in Fig. 2(e). The bright stripes can also be assumed as the phase-shifted domain walls between the neighboring S-CDW domains, which is highlighted by the overlaid CCDW superlattice, as shown in Fig. 2(f). The local electronic states of the bright stripes also exhibit a V-shaped feature, similar to the bright clusters. However, compared with stable local bright clusters, the global bright stripe network is relatively unstable with various dynamic behaviors, including breaking, shrinking, and disappearing, as shown in Fig. S2. The isolated stripe loops can be decomposed from the global bright stripe network, indicating their specific topological stability. According to the global and dynamic characteristics of bright stripes, the emergence of the bright stripe network can be attributed to the preliminary role played by *e-ph* coupling via the interplay with the intralayer electron-electron correlation in the correlated CDW state.

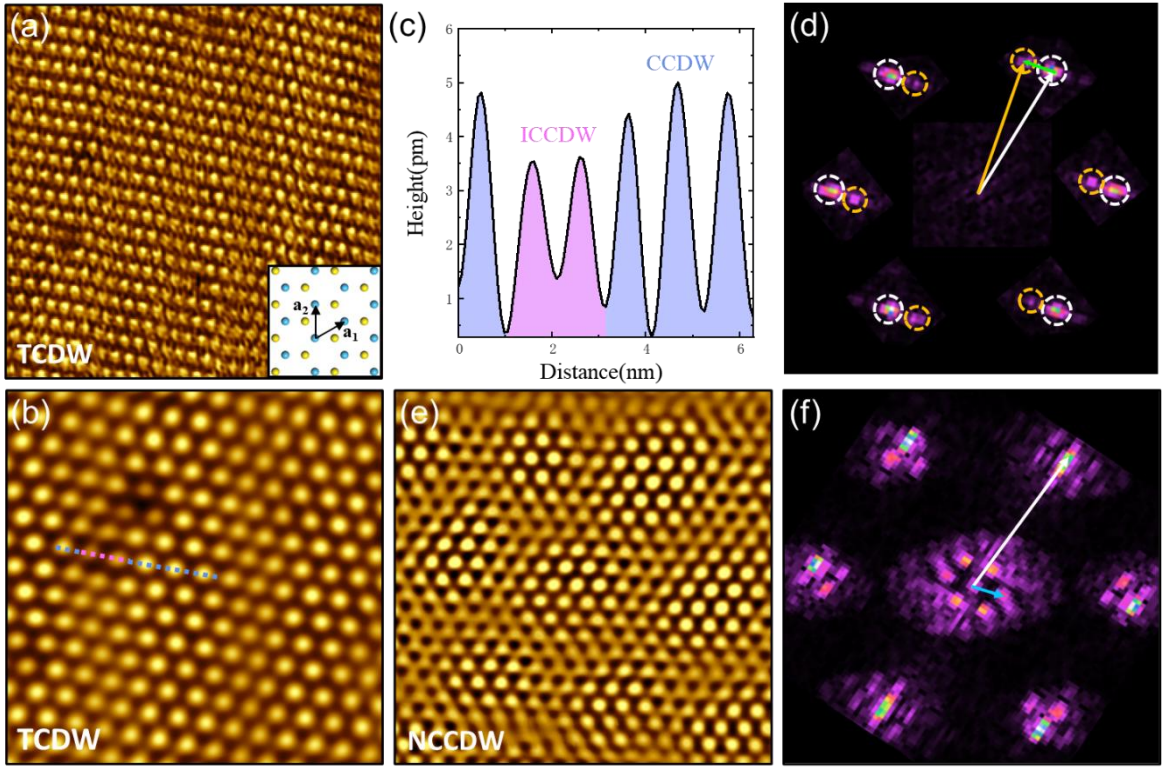

**FIG. 3. STM measurements of the TCDW and NCCDW phases of 1$T$-TaS$_2$ at temperatures above ~230K.** (a) Large-scale STM image of the stripe-like TCDW phase. The stripes are arranged in the high-symmetric directions of the underlying atomic lattice. The inset shows the basis vectors of the underlying atomic lattice. (b) High-resolution STM image of the CCDW and ICCDW stripes in the TCDW phase. (c) Topographic line-profile along the dashed line in (b), showing the rough periods of ICCDW and CCDW stripes. (d) Fast Fourier transform (FFT) pattern of the TCDW phase. The green (white) arrows indicate the periodical stripe-like arrangements of CDW/ICCDW domains (hexagonal arrangements of SDs) in the TCDW phase. (e) High-resolution STM image of the NCCDW phase. (f) FFT of (e). The blue (white) arrows indicate the hexagonal arrangements of CCDW domains (SDs) in the NCCDW phase, respectively. Imaging parameters: 50 × 50 nm$^2$. $V_b$ = 100 mV, $I_t$ = 150 pA (a); 25 × 25 nm$^2$. $V_b$ = 100 mV, $I_t$ = 150 pA (b); 20 × 20 nm$^2$. $V_b$ = 100 mV, $I_t$ = 150 pA (e).

At temperatures above ~230 K, a stripe-like TCDW phase is observed, as shown in Fig. 3(a). The TCDW phase is composed of the bright stripes of enhanced CDW maxima (CCDW), separated by dark stripes of diminished CDW amplitude (ICCDW) as shown in Fig. 3(b). The stripes distribute along the high-symmetry directions of the underlying atomic lattice, showing an exotic spontaneously broken rotational symmetry relative to the low-temperature hexagonal CCDW phase. The Topographic line-profile of Fig. 3(c) is measured along the dashed line in Fig. 3(b), indicating a rough ratio 2:3 of ICCDW and CCDW stripes. The average periodicity of the TCDW phase can be quantitatively determined via the Fast Fourier transform (FFT) of its large-scale STM images, as shown in Fig. 3(d). The stripe-like and hexagonal arrangements of the SDs in the TCDW phase are represented by the satellite and main FFT spots, marked by the small orange and large white circles, respectively, in Fig. 3(d). The averaged periodicity of the TCDW phase can be determined as five SDs by the relative ratio of green and white arrows, i.e., 1:5.

At temperatures above ~280 K, the well-known NCCDW phase emerged from the stripe-like TCDW phase with the reappeared three-fold rotational symmetry, as shown in Fig. 3(e). The NCCDW phase consists of the hexagonal array of CCDW domains separated by the connected ICCDW network. The averaged periodicity of the NCCDW phase can be quantitatively determined by its FFT in Fig. 3(f), where the hexagonal arrangements of the CCDW domains and SDs are represented by the satellite and main FFT spots, respectively. The rough periodicity (2:5) of the CCDW domain superlattice is further determined by the length ratio (1:7) of the blue and white vectors. In this temperature range, the stacking-dependent interlayer interactions completely vanish and *e-ph* coupling gradually plays more considerable role. The nematic TCDW and hexatic NCCDW phases can be explained by the interplay of intralayer electron-electron correlation and *e-ph* coupling.

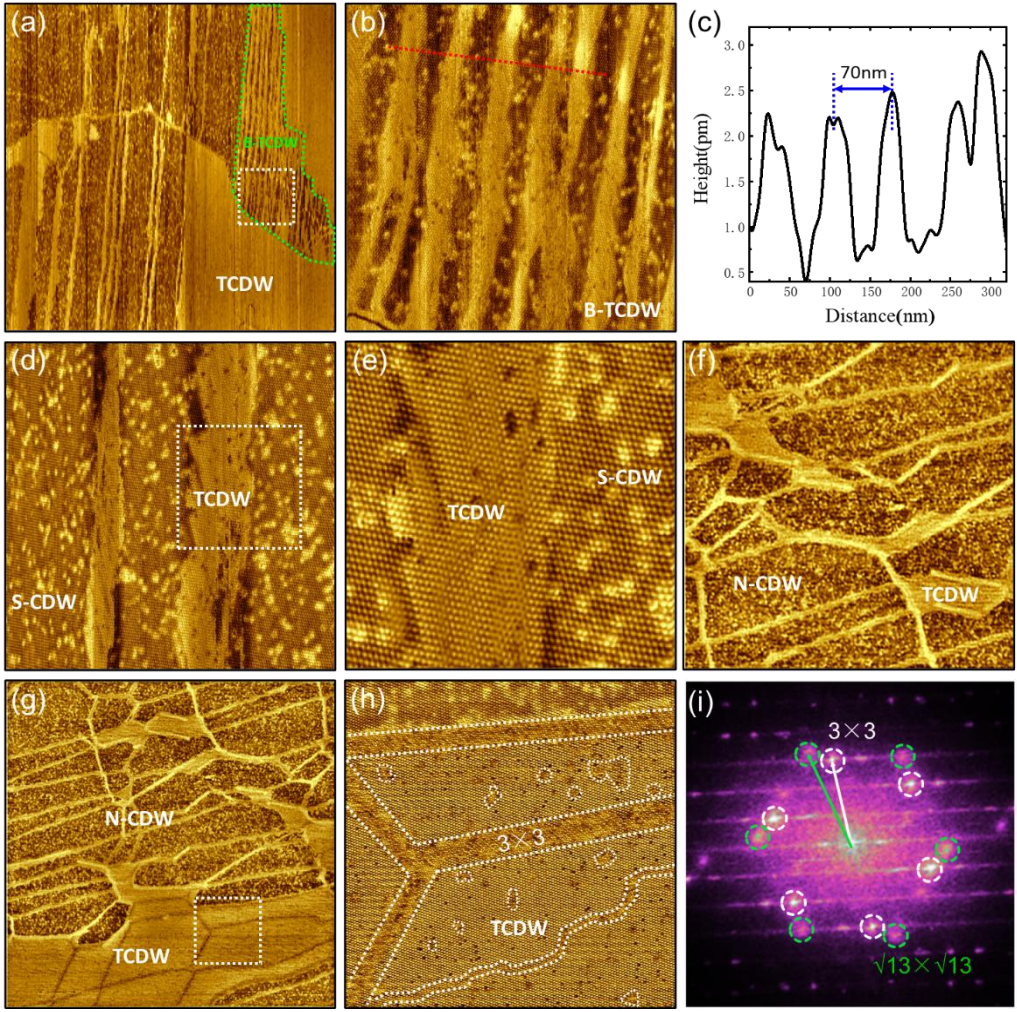

**FIG. 4. Emergent belt-TCDW(B-CDW) and N-CDW electronic states within the TCDW phase during cooling process from 230K.** (a) Mesoscopic-scale STM image of the parallel TCDW belts, which is called the B-TCDW state. The TCDW area that has undergone a phase transition exhibits a striped pattern. The initial stage of the B-TCDW state decomposed from the TCDW phase highlighted by the green dashed line. (b) High-resolution STM image of the B-TCDW state. (c) Topographic line-profile measured along the red line in (b). The rough periodicity and width of the parallel belts are estimated as ~70 nm and ~35 nm, respectively. (d, e) Large-scale and atomic-scale STM images of the TCDW belts; (e) shows a zoomed-in STM image of the area marked by the white dashed square in (d). (f, g) Mesoscopic-scale STM image of the N-CDW state with the residual TCDW regions. (g, h) Mesoscopic-scale and high-resolution STM images of the newly emergent 3 × 3 CDW state within the residual TCDW regions, appearing as the connected dark stripes marked by the dotted lines. (i) FFT pattern of (h). The emergent 3 × 3 and pristine √13 × √13 CDW superlattices are marked by white and green circles, respectively. Imaging parameters: 1.5 × 1.5 um² (a); 300 × 300 nm² (b); 150 × 150 nm² (d); 50 × 50 nm² (e); 500 × 500 nm² (f); 1.0 × 1.0 um² (h); 70 × 70 nm² (i). $V_b$ = 200 mV, $I_t$ = 100 pA (a–i).

Direct cooling from the stripe-like TCDW phase, the transition from the TCDW phase to the CCDW phase are investigated as shown in Figure 4, while the previous work mainly focused on the NCCDW–CCDW transition [39–41]. Figure 4(a) shows a typical STM image of the bright belts that emerged through the split-decompositions process from the TCDW phase, which is named as belt-TCDW (B-TCDW) state. The detailed formation process of the B-TCDW state is shown in Fig. S3. The zoom-in STM image of the B-TCDW region with the dense bright belts (the initial stage of the B-TCDW state) is shown in Fig. 4(b). The S-CDW state with bright clusters reappeared between the bright TCDW belts. The rough periodicity and width of the parallel belts are estimated to be ~70 nm and ~35 nm, respectively, according to the topographic line-profile of Fig. 4(c). Large-scale and atomic-scale STM images of the further decomposed B-TCDW state are shown in Figs. 4(d) and 4(e), respectively. The shrinking widths and rugged edges of the TCDW belts in Fig. 4(d) indicate that the reappearance of bright clusters is caused by the further decomposition of the TCDW belts. The interval CCDW and ICCDW stripes are still observed in the TCDW belts, as shown in Fig. 4(e), suggesting that the two coexisting characteristic lengths of ~6 nm (TCDW) and ~70 nm (B-TCDW) play decisive roles at the atomic-scale and mesoscopic-scale, respectively, during the entire split-decomposing process of the TCDW phase. Although the detailed underlying mechanisms of these characteristic lengths are unclear, we can also attribute them to the delicate competition between the electron correlation and *e-ph* coupling interactions.

At temperatures below ~200 K, a similar N-CDW state reemerged, with a few small residual TCDW regions, as shown in Fig. 4(f). Surprisingly, some unexpected dark stripes are observed within the residual TCDW regions in Fig. 4(g), and are clearly resolved as a newly emergent 3 × 3 CDW state, marked by the white dotted lines in Fig. 4(h). The relative periodicity and orientation of the emergent 3 × 3 CDW and √13 × √13 CDW are shown in Fig. 4(i). Recently, a similar 3 × 3 CDW state has also been observed on the water-covered 1$T$-TaS$_2$ surface, which can be explained by the shifting of phonon softening mode [43]. Here, the emergence of the 3 × 3 CDW can be phenomenally attributed to the modified *e-ph* coupling by the enhancement of electron-electron correlation at the temperature of ~200 K, while the detailed underlying mechanisms need further investigations and will be discussed in our future work.

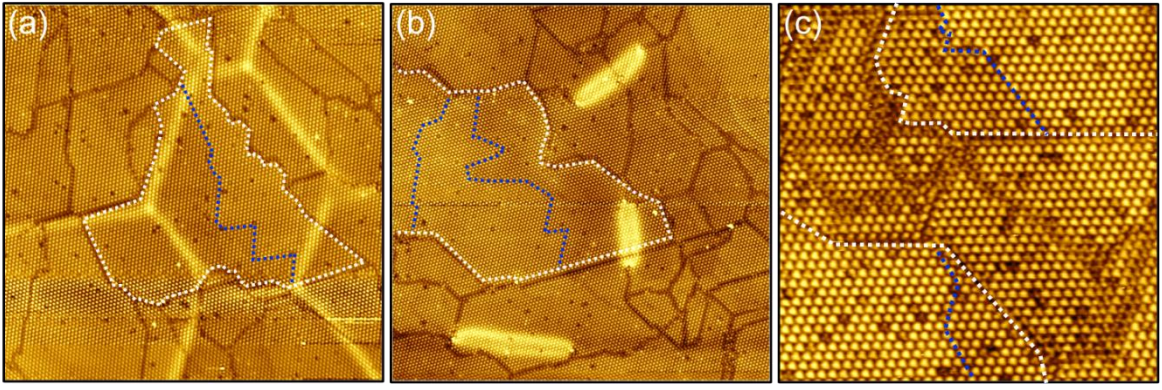

**FIG. 5. Emergent mosaic-CDW (M-CDW) state from the N-CDW state during further cooling process from 160K.** (a, b) Large-scale STM image of the M-CDW state with various stacking orders. The residual bright stripes can still be observed. (c) High-resolution STM image of the M-CDW state. Imaging parameters: 120 × 120 nm$^2$. $V_b$ = 400 mV, $I_t$ =150 pA (a); 120 × 120 nm$^2$. $V_b$ = 400 mV, $I_t$ =150 pA (b); 40 × 40 nm$^2$. $V_b$ = 100 mV, $I_t$ = 150 pA (c).

At temperatures below ~160 K, the N-CDW state completely transforms into the global mosaic CDW (M-CDW) state with a few residual bright stripes, as shown in Figs. 5(a) and 5(b). The global M-CDW state composes of irregularly distributed nano-sized CCDW domains separated by well-defined domain wall textures [46, 47], obviously different from the perfect CCDW state via direct cooling from the NCCDW states [24, 45]. Within the large CCDW domains of the surface layer (marked by white dashed lines), the two apparent heights can be faintly distinguished and separated by the marked blue lines. These different apparent heights in the surface layer sensitively depend on the specific stacking orders of the sublayers, and the domain wall existing in the sublayer(s) can also be visualized within the surface CCDW domains, as indicated by the blue lines in Figs. 5(a)–5(c).

**Discussion**

The competition and/or cooperation of diverse interactions, including stacking-dependent interlayer interactions, intralayer electron-electron correlation, and *e-ph* coupling interactions, are responsible for the various intriguing electronic states with multi-scale features (from atomic-scale to mesoscopic-scale) of 1$T$-TaS$_2$ in the hysteretic CCDW–TCDW phase transition range. The phase transition of CCDW–TCDW is widely considered to be a first-order phase transition [23–25, 34], originating from the significant role played by the *e-ph* coupling of 1$T$-TaS$_2$. Based on the foregoing discussion of various complex interactions, a new Kagome electron lattice can be obtained in 1$T$-TaS$_2$ by further precisely modifying the *e-ph* coupling interactions through the excitation of correlated electrons. In addition, more novel exotic electronic states of 1$T$-TaS$_2$ can be realized with further accurate and effective control of the interlayer and intralayer interactions via intercalation with metallic atoms, electron, or hole doping, etc. The Cu-intercalated 1$T$-TaS$_2$ weakens the interlayer coupling by suppressing the stacking order effect, and enhances the interaction between intralayer SDs, inducing unique modulation of the electronic state [21]. Hole doping with Ti in 1$T$-TaS$_2$ can validly weaken the intralayer electron-electron correlation by reducing the number of intralayer electrons, suppressing the long-range CDW order, and no correlated CDW states were observed at temperatures above 160 K. However, electron doping with W in 1$T$-TaS$_2$ effectively enhances the intralayer electron-electron correlation by increasing the number of intralayer electrons, reproducing the same correlated S-CDW and N-CDW states of pristine 1$T$-TaS$_2$ at low temperatures, which can be considered as a representative electron correlation system. More importantly, our STM observation demonstrates the capacity of Ti- or W-doped 1$T$-TaS$_2$ as a practical method that creates exotic electronic states with finely tuning interlayer interactions and intralayer electron-electron correlation. These results make the 2D materials, as a new category of correlated materials to search for strongly correlated phenomena and exotic electronic states by precisely controlling various interactions.

**Conclusion**

In summary, VT-STM measurements were performed to investigate the spatially hysteretic electronic phase transitions in the correlated CDW state of 1$T$-TaS$_2$. The dynamic behaviors of S-CDW, N-CDW, B-TCDW, and M-CDW states within the CCDW/TCDW phase were detailed studied at the atomic- and mesoscopic-scales through a warming and cooling process. We discuss the influence of three distinct interactions: interlayer interaction, intralayer electron-electron correlation and e-ph coupling interactions on the novel emergent electronic states of 1$T$-TaS$_2$. Deeply understanding the delicate interplay and accurately regulating the various interactions of 1$T$-TaS$_2$ will motivate more research to find the exotic electronic states in the low-dimensional materials.

**Materials and Methods**

  High-quality 1$T$-TaS$_2$ single crystals supplied by HQ Graphene were grown using the chemical vapor transport (CVT) method with iodine as a transport agent. The samples were cleaved at room temperature under ultrahigh vacuum conditions at a base pressure of 2×10$^{-10}$ Torr and quickly transferred to the variable-temperature STM system (PanScan Freedom, RHK, USA). The variable-temperature STM measurements were performed using the liquid-He-free cryocooler technology with an adjustable heating power at the cold head. Chemically etched tungsten tips were used for constant-current STM measurements. The tips were prepared and calibrated on a clean Ag (111) (MaTeck, Germany) surface, which was cleaned by repeated cycles of argon ion sputtering and annealing at 550 K. The STM measurements were mostly performed in the hysteresis temperature range of 1$T$-TaS$_2$, and the investigation of the three phases, i.e., CCDW, TCDW, and NCCDW were investigated via precise regulation of the temperature. STS measurements were also performed at high temperatures (~160 K), which were challenging owing to the thermal-broadening effect. The STM data were analyzed and processed with the Gwyddion software.

**Acknowledgements:** This project is supported by the National Natural Science Foundation of China (NSFC) (No. 21622304, 61674045, 11604063), the Ministry of Science and Technology (MOST) of China (No. 2016YFA0200700), the Strategic Priority Research Program (Chinese Academy of Sciences, CAS) (No. No. XDB30000000). Z. H. Cheng was supported by the Fundamental Research Funds for the Central Universities and the Research Funds of Renmin University of China (No. 21XNLG27).